\documentclass[twocolumn]{aastex631}
\newcommand{\change}[1]{#1}
\graphicspath{ {figures/} }

\begin{document}

\title{The impact of Solar magnetic field configurations on the production of gamma rays at the Solar disk}

\author[0000-0001-6692-6293]{Julien D\"orner}
\author[0000-0002-7388-6581]{Frederic Effenberger}
\author[0000-0002-9151-5127]{Horst Fichtner}
\affiliation{
    Institut f\"ur Theoretische Physik IV, 
    Fakult\"at f\"ur Physik und Astronomie, \\
    Ruhr-Universit\"at Bochum, 
    Universit\"atsstra\ss e 150, 44780 Bochum, Germany
}
\affiliation{Ruhr Astroparticle and Plasma Physics Center (RAPP Center), 44780 Bochum, Germany}

\author[0000-0002-1748-7367]{Julia Becker Tjus}

\affiliation{
    Institut f\"ur Theoretische Physik IV, 
    Fakult\"at f\"ur Physik und Astronomie, \\
    Ruhr-Universit\"at Bochum, 
    Universit\"atsstra\ss e 150, 44780 Bochum, Germany
}
\affiliation{Ruhr Astroparticle and Plasma Physics Center (RAPP Center), 44780 Bochum, Germany}

\affiliation{Department of Space, Earth and Environment, Chalmers University of Technology, SE-412 96 Gothenburg, Sweden}

\author[0000-0002-9672-3873]{Meng Jin}
\affiliation{Lockheed Martin Solar and Astrophysics Laboratory, 3251 Hanover Street, Palo Alto, CA 94306, USA}

\author[0000-0001-8794-3420]{Wei Liu}
\affiliation{Bay Area Environmental Research Institute, NASA Research Park, Building 18, Mailstop 18-4, Moffett Field, CA 94035-0001, USA}
\affiliation{Lockheed Martin Solar and Astrophysics Laboratory, 3251 Hanover Street, Palo Alto, CA 94306, USA}

\author[0000-0002-2670-8942]{Vahe' Petrosian}
\affiliation{Department of Physics and KIPAC, Stanford University, Stanford, CA 94305, USA}
\affiliation{Department of Applied Physics, Stanford University, Stanford, CA 94305, USA}

\begin{abstract}
    The Sun produces a steady signal of high-energy gamma rays through interactions of Galactic cosmic rays (GCRs) with its atmosphere. Observations with Fermi-LAT and HAWC have revealed a gamma-ray flux significantly higher than early theoretical predictions, with unexpected temporal and spectral features that suggest a crucial role of the solar magnetic field. In this work, we model GCR-induced gamma-ray emission at the solar disk using the CRPropa framework with realistic hadronic interactions, chromospheric density profiles, and several magnetic field configurations over the solar cycle. 
    This allows us to quantify the gamma-ray emission of the entire solar disk for different phases of the solar activity cycle and we present, for the first time, maps of the production locations of gamma rays on the solar surface. We consider both mono-energetic and realistic power-law injection spectra in a simplified dipole-quadrupole-current-sheet model and potential-field source surface (PFSS) extrapolations for Carrington rotations during solar maximum and minimum. 
    Our results show that magnetic mirroring and large-scale field topology strongly affect the spectral shape and spatial distribution of the emission, with slightly enhanced fluxes predicted at solar minimum. While our simulated baseline fluxes remain below observations, additional effects, such as heavier nuclei, Parker-field mirroring, and deeper atmospheric interactions, could result in further enhancements of fluxes closer to observational values.
    Hadronic interactions do not only produce gamma rays but also neutrinos. We estimate the expected neutrino flux from the Sun based on our predictions. We find that the expected flux is slightly below current upper limits from IceCube.
\end{abstract}


\section{Introduction}
The dominant high-energy cosmic ray component in the heliosphere is of Galactic origin. These Galactic Cosmic Rays (GCRs) are not only useful messengers from distant astrophysical systems \citep[e.g.,][]{Becker-Tjus-Merten-2020}, but they can also be used to infer information about the large-scale structure of the heliosphere \citep[e.g.,][]{Pogorelov-etal-2017} as well as about the turbulence in the solar wind \citep[e.g.,][]{Oughton-Engelbrecht-2021, Fraternale-etal-2022}. 

Even in the innermost heliosphere, high-energetic GCRs are important agents regarding the solar magnetic field. The deflection of GCRs in the latter produces the so-called solar cosmic-ray shadow \citep[][and references therein]{Becker-Tjus-etal-2020, Aartsen-etal-2021}, which varies with solar activity and can, therefore, be used to infer information about the solar coronal magnetic field \citep[e.g.,][]{Alfaro-etal-2024, Ng-etal-2025}. The GCRs also interact with the atmosphere and the interior of the Sun. \citet{Seckel-etal-1991} were the first to estimate the fluxes of gamma-rays, neutrinos, antiprotons, neutrons, and antineutrons resulting from cascades that are initiated by the interactions of GCRs with the solar atmosphere. Aside from protons and heavier ions, GCR electrons and positrons can also produce gamma-rays via inverse Compton scattering of solar photons, as well as extreme ultraviolet (EUV) to hard X-ray emission via synchrotron radiation in the solar magnetic field. The transport of GCR electrons from 1~AU to the Sun and their subsequent synchrotron emission were recently modeled \citep{Petrosian-etal-2023, Orlando-etal-2023}.
The steady flux of gamma-rays produced near the Sun as a result of GCR proton interactions is the primary interest in the present study.

Gamma rays produced by the Sun in the GeV range are also associated with transient solar flare events. The maximum energy of flare-produced gamma rays is about 4~GeV \citep[e.g.][]{Ackermann-etal-2014}, which is lower than most of the energy range discussed in the following, reaching well into the TeV regime. Regarding the steady gamma-ray flux of the Sun, one distinguishes two components of the GCR-induced gamma-ray emission, namely the disk and the halo component. The first one comes from the solar disk and was theoretically studied by \citet{Seckel-etal-1991} and detected first in EGRET data \citep{Orlando-Strong-2008}. It results predominantly from GCR protons, which follow magnetic flux tubes in the solar atmosphere. When these particles interact with the solar atmosphere, they produce, via cascades, gamma-rays from the direction of the solar disk. The second, so-called halo component, results from inverse Compton scattering of cosmic-ray electrons in the extended solar atmosphere, quantified and validated by \citet[][see also the review in \citet{Orlando-Strong-2021}]{Orlando-Strong-2008} as far as possible with the EGRET data available at that time \citep[see also][]{Moskalenko-etal-2006}. The predictions for the halo component and the corresponding EGRET observations have been confirmed with measurements made with the Large Area Telescope of the \emph{Fermi} mission \citep[\emph{Fermi}-LAT,][]{Abdo-etal-2011}. The disk component was detected with \emph{Fermi} as well, however, at a flux level about a factor of six higher \citep{Nisa-etal-2019} than predicted by \citet{Seckel-etal-1991}. These findings have been confirmed at lower energies in \citet{Ng-etal-2016} by extending the analysis of \emph{Fermi}-LAT data from 1--10~GeV \citep{Abdo-etal-2011} to the interval 1--100 GeV. With the High Altitude Water Cherenkov (HAWC) Observatory \citep{DeYoung-2012}, solar disk gamma rays with higher energies in a low TeV range have been observed \citep{Albert-etal-2023}. 


\citet{Ng-etal-2016} identified a time-dependence of the flux of the disk component by a factor of two to three in the energy interval $1-10$~GeV, confirmed by \citet{Linden-etal-2022}. This variation is present up to about 30~MeV, anti-correlated to solar activity, and cannot be explained exclusively with the corresponding modulation of GCRs, which amounts to $\sim$15\% for 10~GeV protons and less for higher energies. So, as for the cosmic-ray shadow, the solar magnetic field must be expected to play a significant role: The field structure determines the trajectories of the GCRs in the solar atmosphere. First, it determines which particles initiate cascades that eventually lead to the formation of the disk component of gamma rays. And, second, the field also determines the temporal variation of the gamma-ray signal. Such variation with solar activity is also present in the HAWC data \citep{Albert-etal-2023}. 

Furthermore, \citet{Tang-etal-2018} reported an unexpected dip in the solar gamma-ray spectrum between $30-50$~GeV, which has recently been attributed to an energy cutoff in the turbulence-induced migration of GCRs into closed magnetic arcades \citep{Puzzoni-etal-2024}. While the confirmation of this needs further measurements, another finding remains, so far, unexplained. \citet{Linden-etal-2018} reported that the gamma-ray emission is not point-like and isotropic \citep{Nisa-etal-2019} but exhibits a nearly constant polar component up to 100~GeV and an equatorial component that is brightest during solar minimum with energies above 200 GeV but not in the TeV-regime \citep{Albert-etal-2018, Bartoli-etal-2019}. Finally, the gamma-ray flux is characterized by power laws in energy, namely $E^{-2.2}$ below 200 GeV \citep[][and references therein]{Linden-etal-2022}, steepening to $E^{-3.6}$ above 500~GeV \citep{Albert-etal-2023}.

Following the original idea by \citet{Seckel-etal-1991}, recently \citet{Li-etal-2024} have attempted to identify the magnetic field structures at the solar surface that are crucial for reflecting hadronic GCRs and have modeled the corresponding gamma-ray emission. To this end, the authors considered a solar surface magnetic field consisting of vertical flux tubes, representing those that form the so-called network elements, and of vertical flux sheets, representing those at the so-called down-flow lanes between granules. This way, the discrepancy between simulations and observations could be somewhat reduced, but it remains at a factor of two to four. While this model is quite advanced on a small scale, it assumes a spherically symmetric solar atmosphere on a large scale and, thus, does not take into account the large-scale coronal and interplanetary magnetic field structure. In the context of gamma-ray production, the coronal field very close to the Sun was taken into account by \citet{Hudson-MacKinnon-2020} using an MHD model and by \citet{Li-Ng-etal-2024} employing the potential field source surface (PFSS) approach, again resulting in too low simulated fluxes. 

All of the above models did not consider a small-scale turbulent component of the large-scale solar magnetic field. This was done recently by \citet{Puzzoni-etal-2024}, focusing on closed turbulent magnetic field arcades in the low photosphere on the gamma ray flux. This study was complemented by \citet{Puzzoni-etal-2025} with modeling a synthetic large-scale magnetic field with static, laminar open field lines in the chromosphere that are increasingly braided near the solar surface. A key finding of these studies is that the tension between the modelled and the observed gamma ray flux from the solar disk reduces with increasing turbulence level. In extension of their earlier work, \citet{Li-etal-2025} also included the effect of turbulence. Interestingly, the Alfv\'enic turbulence considered therein was found to reduce rather than increase the modelled flux. 

Extending the scale to a few solar radii, \citet{Hutchinson-etal-2022} considered the interplanetary Parker field and the turbulence-induced scattering of GCRs within, however, only in the context of solar flares and not yet computing gamma ray fluxes.

With the present paper, we execute the next step in the modeling of the GCR-induced gamma-ray production in the solar atmosphere by computing the trajectories of energetic protons inside the so-called source surface in different global PFSS configurations corresponding to different Carrington rotations (CRs) magnetic maps. This allows us to quantify the gamma-ray emission of the entire solar disk for different phases of the solar activity cycle. The paper is structured as follows: in Section~2, the modeling framework is described in detail. In Section~3, the results are presented and critically discussed. The final section~4 contains the conclusion and provides an outlook on future work. 

\section{Modelling Framework}
We model the gamma-ray emission with CRPropa 3.2 \citep{CRPropa3.2}, a publicly available tool for the propagation and interaction of charged particles in magnetic fields. While the code was originally written for ultra high energy cosmic rays (UHECRs) \citep{CRPropa3}, an extension for lower energies to cover GCRs has been introduced later on \citep{CRPropa3.1}. The CRPropa framework allows modeling the transport behavior of energetic particles by either solving the Lorentz-Newton equation (so-called ballistic propagation) or by applying a set of stochastic differential equations (SDEs) to solve the GCR transport equation. 

\begin{figure}
    \centering
    \includegraphics[width=\linewidth]{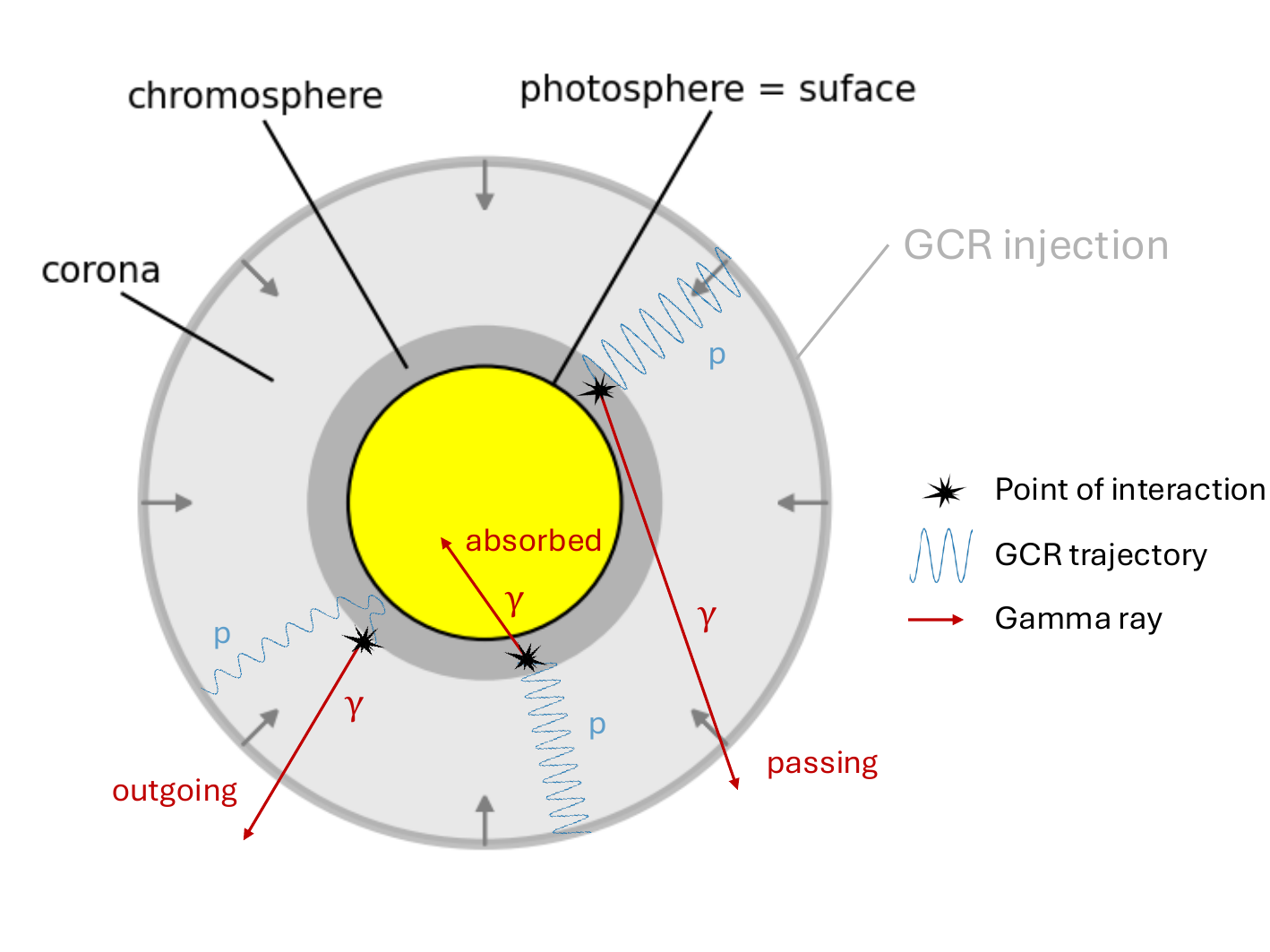}
    \caption{Schematic illustration of the simulation setup.}
    \label{fig:schema}
\end{figure}

For the application of GCRs in the atmosphere of the Sun, we start by using the ballistic approach, in which transport effects like the mirroring of particle trajectories due to gradients in the magnetic field strength can be modeled. For further studies, a direct comparison between the diffusive and the ballistic approach would be useful, but this is beyond the scope of this paper. A schematic illustration of the simulation setup is shown in Fig.~\ref{fig:schema}.

\subsection{Magnetic field configuration}
\begin{figure*}[ht]
    \centering
    \includegraphics[width=\textwidth]{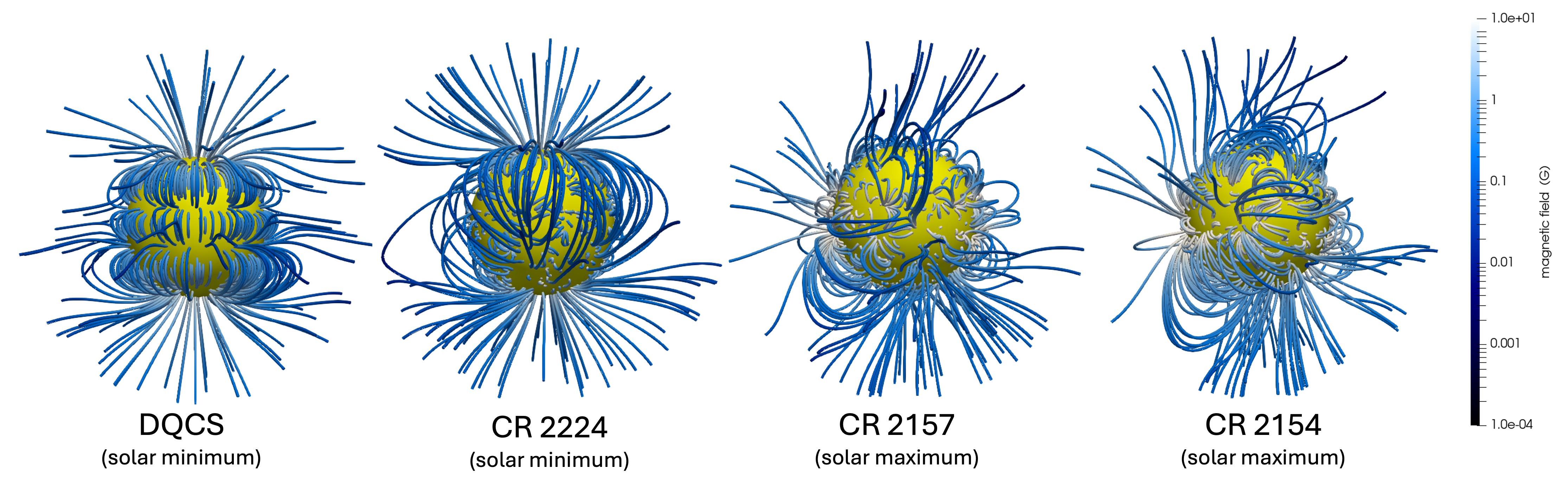}
    \caption{Magnetic field configuration in the DQCS and the PFSS model for three different Carrington Rotations.}
    \label{fig:fields}
\end{figure*}

In the simulation, different magnetic field configurations can be used. The baseline model considered here is the Dipole + Quadrupole + Current Sheet (DQCS) model introduced by \cite{Banaszkiewicz_DQCS}. This model consists of a dipole and quadrupole moment and a contribution from the current sheet in the Sun's equatorial plane. It fits the solar magnetic field quite well in solar minimum conditions. A set of field lines is shown in Fig.\ \ref{fig:fields}. In CRPropa, the field is evaluated at each position based on the analytical description. 

To evaluate the influence of the magnetic field configuration, we compare the results of the baseline field with the potential field source-surface (PFSS) model \citep{Schatten69, Altschuler69} evaluated at different times. 
We calculate the PFSS field using the \textit{pfss} Python package from Antony Yeates\footnote{\url{https://github.com/antyeates1983/pfss}} on a spherical grid containing 60 bins in $\log(r)$, 180 bins in $\cos\theta$, and 360 bins in $\phi$. The outer surface, at which the magnetic field is assumed to be purely radial, is set to $R_{ss} = 2.5 R_\odot$. In between the grid points the magnetic field is evaluated by tri-linear interpolation in spherical coordinates, although it is known that the interpolation might affect the results \citep{Schlegel20}. 

The input of the PFSS field calculation is a magnetic map reconstructed from a 27-day solar rotation (Carrington Rotation, CR), i.e. synoptic or diachronic magnetic map. We use the synoptic maps measured with the Helioseismic and Magnetic Imager (HMI) \citep{Schou-etal-2012} for the CR 2154, 2157, and 2224. The first two CR describe the active phase of the Sun, while the last one is close to the solar minimum. An illustration of the field line configuration in the different CRs is shown in Fig.\ \ref{fig:fields}. 

\subsection{Hadronic Interaction}
The gamma rays observed from the Sun are produced in hadronic interactions of GCRs in the solar atmosphere. The dominant interaction channel is the production of neutral pions $p+p\rightarrow \pi^0 \rightarrow \gamma\gamma$. We model the interactions of GCRs with the Hadronic Interaction plug-in\footnote{\url{https://gitlab.ruhr-uni-bochum.de/doernjkj/hadronic-interaction-in-crpropa}} for CRPropa presented by \cite{Doerner25}. 
\change{The module calculates the interaction probability in each propagation step, based on the inelastic cross section by \cite{Kafexhiu-etal-2014} and interactions are performed in a Monte-Carlo method. In the case of an interaction, the energy of all secondaries are sampled from the differential cross-section for a given species. In this work we apply the model} from \cite{ODDK22, ODDK23} as it also includes interactions at the $\sim$GeV scale. All interactions are treated as catastrophic losses, and the interacting GCR is removed from the simulation after the interaction. 

In this CRPropa module, it is assumed that all secondary particles are created in the propagation direction of the primary GCR. This is based on the assumption of ultra-relativistic particles within CRPropa. \citet{Griffith-etal-2025} introduce the effect of the opening angle in the interaction, which can be relevant for proton energies $E_p \lesssim 10 \, \mathrm{GeV}$. In this work, we focus on gamma rays with $E_\gamma \gtrsim 1\, \mathrm{GeV}$, which are predominantly created by protons with $E_p \gtrsim 10 \, \mathrm{GeV}$. Therefore, we do not expect a strong impact of this assumption.

\begin{figure}[t]
    \centering
    \includegraphics[width=\linewidth]{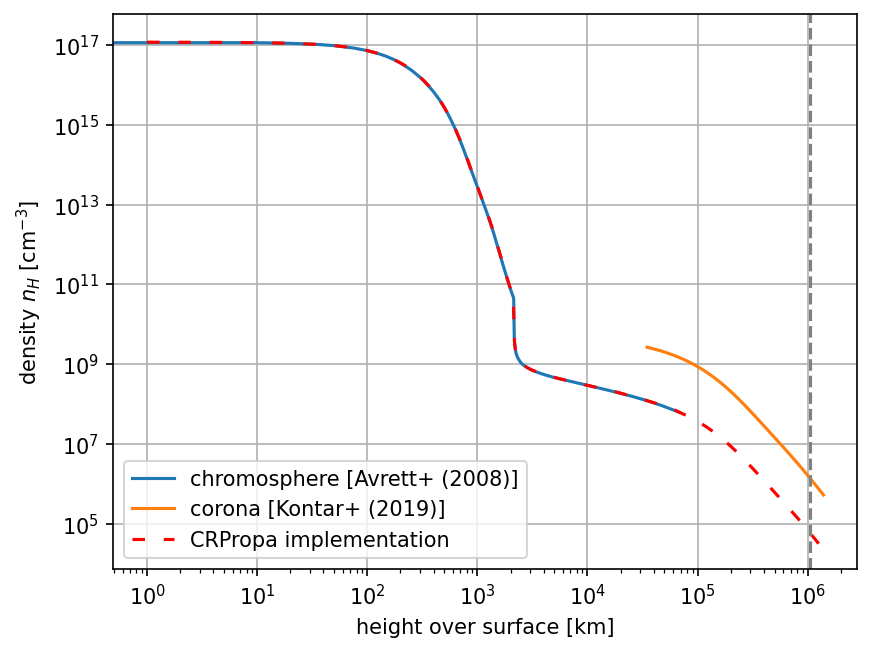}
    \caption{Density profile for the chromosphere and the corona. The dashed grey line indicates the starting position of the GCRs.}
    \label{fig:density}
\end{figure}

Besides the cross-section, the other input needed is the target density.  
We use the chromosphere model from \cite{Avrett08}. To extend the density profile to the source surface at $r = 2.5 \, R_\odot$, we follow the radial slope from \cite{Kontar19}, but apply a lower normalization.
The radial dependence of both models is shown in Fig.\ \ref{fig:density} as a function of the height above the solar surface. 

\subsection{CRPropa setup}
The simulation volume is restricted by two spheres. The inner boundary is set to $r = 1 R_\odot$, referring to the solar surface. The outer boundary is set to $r_\mathrm{inj} = 2.5 R_\odot$, which is the starting position of the particles. 
We distribute the starting positions homogeneously around the sphere, with an isotropic pitch angle distribution. Only those particles with an ingoing direction are taken into account. The deflection of the particles within the solar magnetic field is calculated with the Boris push method implemented in CRPropa. 
For simplicity, we only consider protons as primary GCRs. 

For the energy of the primary GCRs, we consider two scenarios. First, we simulate mono-energetic with a starting energy of $E = 10^0, 10^2$ or $10^4$ GeV. This allows for an easier comparison of the energy dependence of the mirroring and other transport-related effects.

In the second case, we consider a more realistic energy distribution. Here, we simulate a flat proton spectrum
\begin{equation}
    \left.\frac{\mathrm{d}N}{\mathrm{d}E}\right|_\mathrm{sim}(E) = \frac{N_\mathrm{sim}}{\ln(E_\mathrm{max}/E_\mathrm{min})} \, E^{-1}
\end{equation} 
in the energy range from $E_\mathrm{min} = 1\, \mathrm{GeV}$ to $E_\mathrm{max} = 1 \, \mathrm{PeV}$. This flat energy distribution allows equal statistics in each logarithmic energy bin. 
After the simulation, the source spectrum is reweighted to match the Local Interstellar Spectrum (LIS), which we parametrize as 
\begin{equation}
        J_\mathrm{LIS}(E) = J_0 \ E_\mathrm{GeV}^{1.12} \, \beta^{-2}  \left( \frac{E_\mathrm{GeV} + 0.67}{1.67} \right)^{-3.93} \ . \label{eq:LIS}
\end{equation}
Here, $E_\mathrm{GeV}$ describes the kinetic energy in GeV units and $\beta$ is the particle velocity in units of the speed of light. The normalization is $J_0 = 1.08 \cdot 10^{-2}\pi \, \mathrm{GeV}^{-1} \, \mathrm{m}^{-2} \, \mathrm{s}^{-1}$. 
Each observed particle is weighted by a factor 
\begin{equation}
    w_i = \frac{J_\mathrm{LIS}(E_0^i)}{\left. \frac{\mathrm{d}N}{\mathrm{d}E}\right|_\mathrm{sim}(E_0^i)} \, \left(\frac{r_\mathrm{inj}}{d} \right)^2\quad,  \label{eq:weights}
\end{equation}
depending on the energy $E_0^i$ at the source and the distance $d$ between Sun and Earth. 

\begin{figure}[t]
    \centering
    \includegraphics[width=\linewidth]{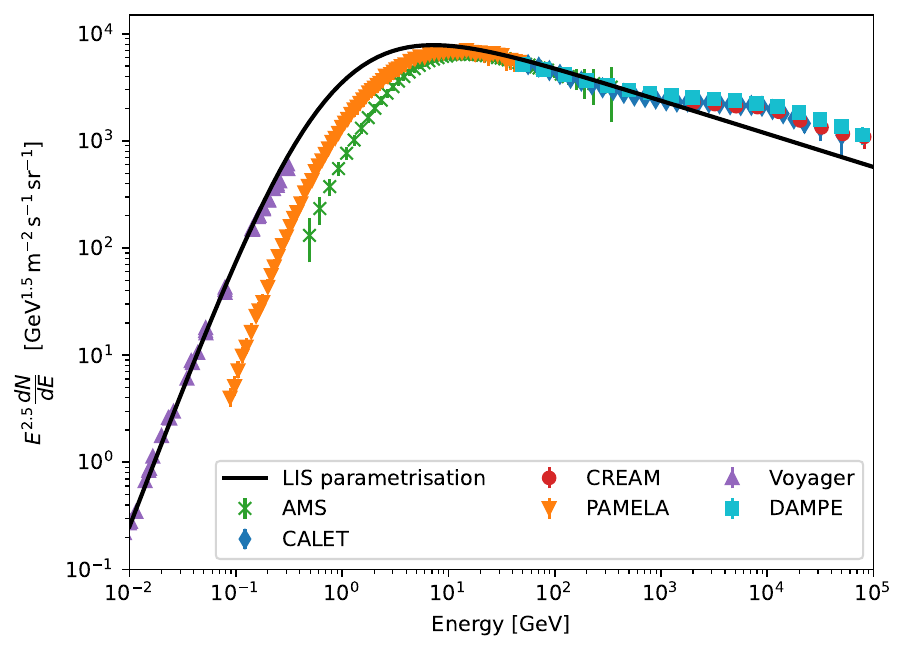}
    \caption{Parametrisation of the local interstellar spectrum with the data measured by AMS-02 \citep{AMS}, CALET \citep{Calet}, CREAM \citep{Cream}, Pamela \citep{Pamela}, Voyager \citep{Voyager13, Voyager16, Voyager19} and DAMPE \citep{Dampe}.}
    \label{fig:LIS}
\end{figure}

In Fig.\ \ref{fig:LIS} the parametrization and current observation of the GCR proton spectrum are shown. 

\subsection{Filtering of gamma-ray directions} \label{ssec:directions}
The production of gamma rays is dominantly in the direction of the parent proton due to the high Lorentz boost. As many of the incoming GCR protons are travelling towards the Sun, most gamma rays are absorbed by the solar disk and can not be observed from the Earth. \change{The absorption of gamma-rays is not taken into account within the CRPropa simulation, but can be treated in the post-processing.} To account for this effect in our simulation, we perform a filtering of the produced gamma rays in the post-processing. Here, we differentiate between three modes illustrated in Fig.~\ref{fig:schema}. For further analysis, we construct three datasets, where every selection includes all particles selected before:
\begin{enumerate}
    \item In the first selection of gamma rays (\textbf{outgoing}), all particles with an outgoing momentum are considered. Those gamma rays can only be produced by GCR protons that have undergone magnetic mirroring before.

    \item In the second selection (\textbf{passing}), we add all gamma rays with an ingoing momentum that will not hit the solar surface (labeled passing in Fig.~\ref{fig:schema}). Those gamma rays are produced by GCR protons with a nearly perpendicular pitch-angle, but the point of interaction is far enough away from the solar surface. Here, we note that the number of particles added in this selection is rather small, as the density drops significantly for larger heights over the surface. \change{Nearly all results shown later on, are the same for the \textbf{outgoing} and \textbf{passing} selections. We refer the reader to focus on the \textbf{passing} results.}

    \item In the third selection (\textbf{all}), all particles are collected. Here, we do not differentiate between those gamma rays that can be observed from Earth and not. This selection has the highest statistic, as most photons are produced in the direction of the solar disk. 
\end{enumerate}

\section{Mono Energetic simulation} \label{sec:monoE}

As a first step, we analyze simulations with a mono-energetic injection to investigate the impact of the energy on several observables, starting with the production height above the surface (Sec.~\ref{ssec:height}), \change{the trapping time of GCRs (Sec.~\ref{ssec:time}),} the initial pitch-angle distribution leading to gamma-ray production (Sec.~\ref{ssec:mu0}) and their spatial distribution on the solar surface (Sec.~\ref{ssec:maps}).

\subsection{Production height of gamma rays} \label{ssec:height}
We test the production height of gamma rays to identify the relevance of the magnetic field structure. To better estimate the influence depending on the initial particle energy, we perform a set of simulations with a fixed starting energy  $E = 1 \, \mathrm{GeV}$, $E = 10^2 \, \mathrm{GeV}$, and $E = 10^4 \, \mathrm{GeV}$.

\begin{figure*}[htb]
    \centering
    \includegraphics[width=\linewidth]{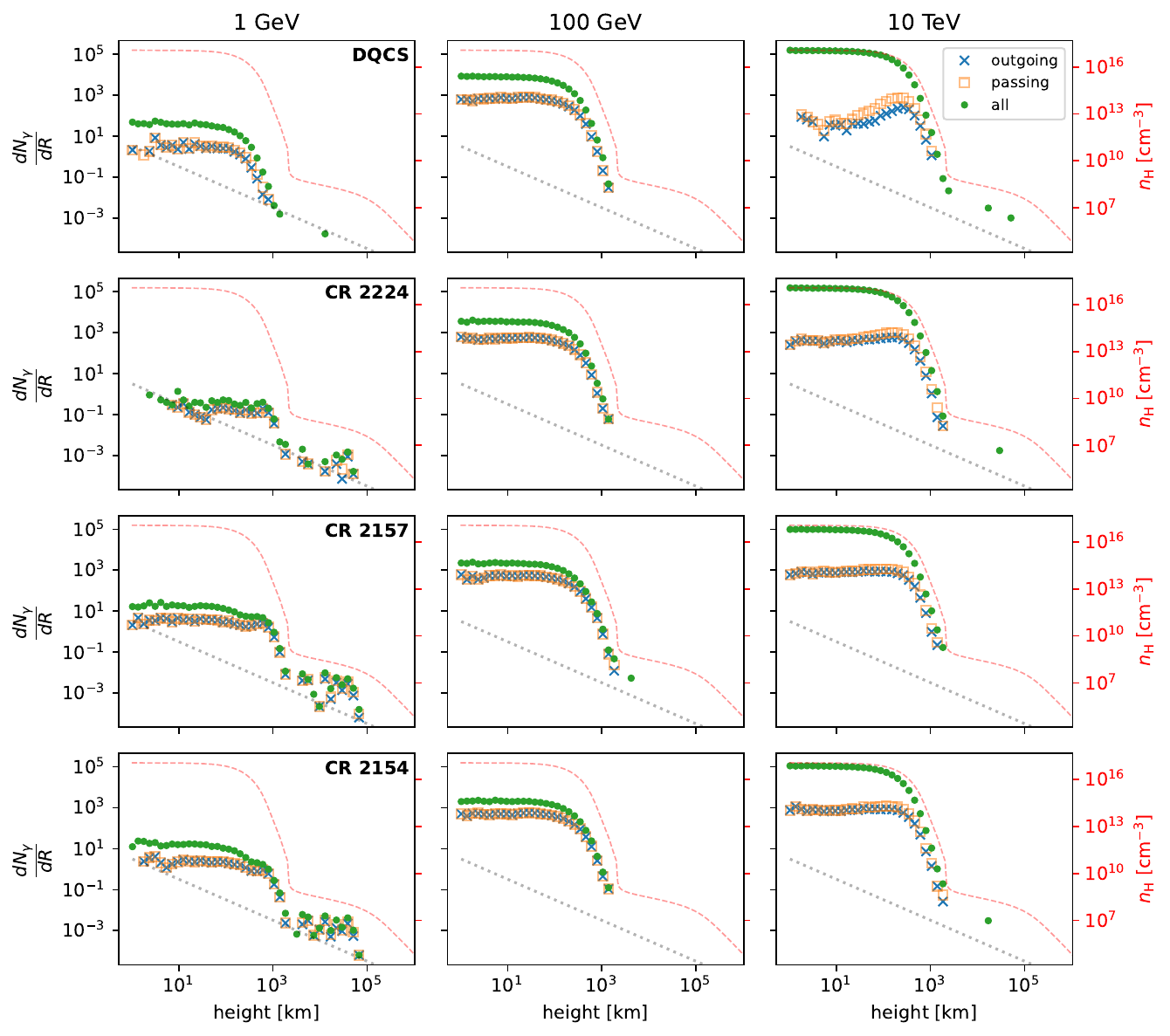}
    \caption{Production height of the gamma rays over the surface in different magnetic field configurations. \change{The row denotes the magnetic field configuration and the column the initial particle energy.} The dotted line shows the minimal resolvable gamma ray density with 1 photon per spatial bin.}
    \label{fig:height}
\end{figure*}

In Fig.\ \ref{fig:height}, the distance of the produced gamma rays to the solar surface is shown. The distribution is shown for all directional filters discussed in Sec.~\ref{ssec:directions}. For a better comparison, the radial density profile is shown as a red dashed line. The minimal resolvable statistic with one observed gamma ray in a radial bin is shown as the gray dotted line. 

The distribution of all produced gamma rays follows directly from the gas distribution. Only differences in the total number of photons depending on the primary energy can be observed. Two different effects lead to the change in the absolute normalization \change{assuming the same column density}. On the one hand, the inelastic proton-proton cross section is energy dependent. This effect is important for the 1 GeV particles, as their energy is close to the kinematic threshold of the interaction. At higher energies, the cross section increases only slowly \citep{Doerner25}. On the other hand, the gamma-ray multiplicity increases for higher energies, leading to more photons in total. \change{Changes in the absolute normalization due to solar modulation have not been taken into account for this simulation.}

The outgoing and passing event selection follows the same radial trend. Only at the highest energies can some deviations be observed. In the more ordered magnetic field configurations (Carrington rotation 2224 and the DQCS model), the gamma-ray production is enhanced between $10^2$ and $10^3$ km above the surface. This effect is stronger in the DQCS magnetic field model, as it is more regular than the solar minimum PFSS model. For the magnetic field configurations at the solar maximum, the field structure is more irregular, and the depth at which a particle is mirrored is more dependent on the starting position. 

\subsection{Trapping of cosmic rays} \label{ssec:time}
\change{
The total column density, which the GCRs transverse depends on the efficiency of trapping the particles in horizontal magnetic field structures. To investigate this effect in the given field configurations, we calculate the time between the injection of GCRs at the outer boundary and the interaction in the solar atmosphere.}

\begin{figure}
    \centering
    \includegraphics[width=\linewidth]{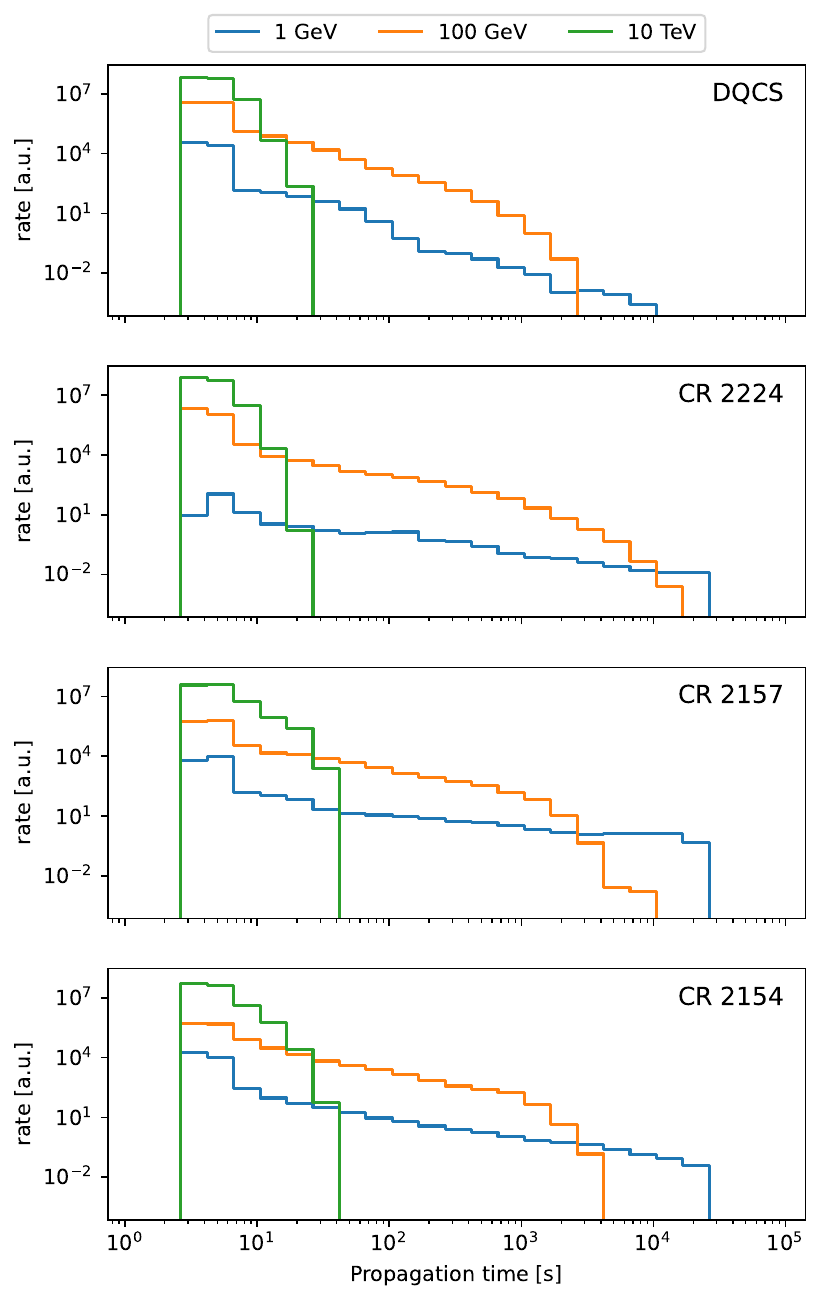}
    \caption{\change{Production rate of secondary gamma-rays dependent on the propagation time of their parent GCRs. The time is taken from their starting surface at $r = 2.5 R_\odot$ until interacting in the solar atmosphere. Here, all produced gamma-rays are shown.}}
    \label{fig:time}
\end{figure}

\change{In Fig.~\ref{fig:time} the emission rate of secondary gamma-rays is shown. In all magnetic field configurations clear differences between the lower ($E = 1$ GeV or $E = 10^2$ GeV) and high ($E = 10^4$ GeV) can be seen. Only GCRs with lower energy can enter loops of closed field lines and are efficiently trapped in the horizontal field. This increases the total confinement significantly and increases the production rate at later times ($t > 10^2$ s).
}

\change{At the highest energy the difference in the field geometry between the more ordered solar minimum cases (DQCS or CR 2224) and the chaotic solar maximum cases (CR 2154 or 2157) can be seen. The more irregular structure at the solar maximum allows even the high energetic GCRs to enter closed field lines. This leads to slightly higher confinement times, but is still less than in the lower energetic case.} 

\change{The horizontal structure of these loops can explain the difference between the absolute normalization in the \textbf{all} and \textbf{outgoing/passing} event selection as seen in Fig.~\ref{fig:height}.}

\subsection{Initial pitch-angle distribution} \label{ssec:mu0}

The distance that a particle can travel in an increasing magnetic field before it is mirrored depends on the initial pitch angle. This effect has been analyzed for GCRs propagating from the orbit of the Earth to the outer boundary of the solar atmosphere by \cite{Hutchinson-etal-2022}. Particles with a nearly perpendicular initial pitch angle will be mirrored earlier and cannot reach the lower layers of the chromosphere, where gamma rays can be produced. 

\begin{figure*}[htb]
    \centering
    \includegraphics[width=\linewidth]{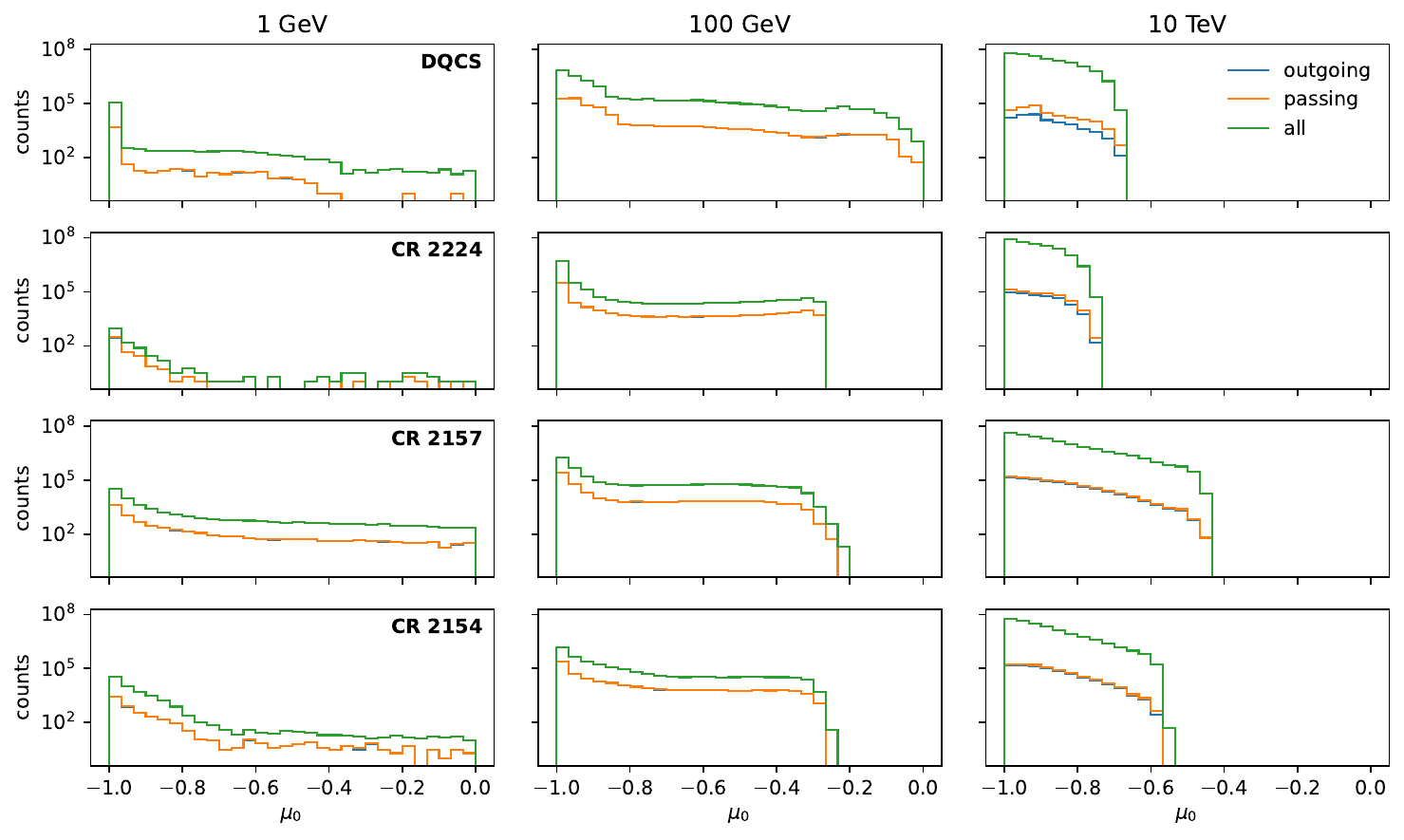}
    \caption{Pitch angle distribution of the initial GCRs leading to the production of gamma-rays. The row denotes the magnetic field model and the column the injected GCR energy.}
    \label{fig:mu0}
\end{figure*}

Fig.~\ref{fig:mu0} shows the distribution of the initial pitch-angle $\mu_0 = \cos(\theta_0)$ from those GCRs that interact and produce gamma rays. All distributions show the expected higher numbers of gamma-ray production for pitch angles in the forward direction. In the intermediate and high energies ($E = 100$ GeV and $E = 10$ TeV), the lower pitch-angles are mirrored before reaching the chromosphere, and no gamma-rays are produced. Only the DQCS model predicts some gamma-ray production for nearly perpendicular starting particles at intermediate energies. This is directly related to the current sheet that is included in this field model. It acts as a highway for the particles, and the gradient in the field is weaker compared to other regions. Therefore, the mirroring happens deeper within the chromosphere, and gamma rays can be produced. 

The critical pitch-angle at which the particles are mirrored before reaching the lower layers of the chromosphere depends on the energy and the magnetic field configuration. In general, a higher energy requires a more forward-directed initial pitch angle. The impact of the magnetic field configuration is less pronounced in the intermediate energy ($100 \, \mathrm{GeV}$) compared to the higher energy. Here, the more ordered field for the solar minimum (CR 2224) and the DQCS field have a narrower pitch angle range, leading to gamma-ray production. This implies the presence of localized structures in the more active field geometries that help the GCRs to reach the lower layers of the atmosphere. 

\subsection{Distribution of the gamma ray production on the solar surface} \label{ssec:maps}

The most direct impact of the magnetic field structure on gamma-ray observables is the position of produced gamma-rays projected onto the solar surface. We calculate the production position in spherical coordinates and histogram the positions with an HEALPix schema \citep{Healpix}. 

\begin{figure*}[hp]
    \centering
    \includegraphics[width=\textwidth]{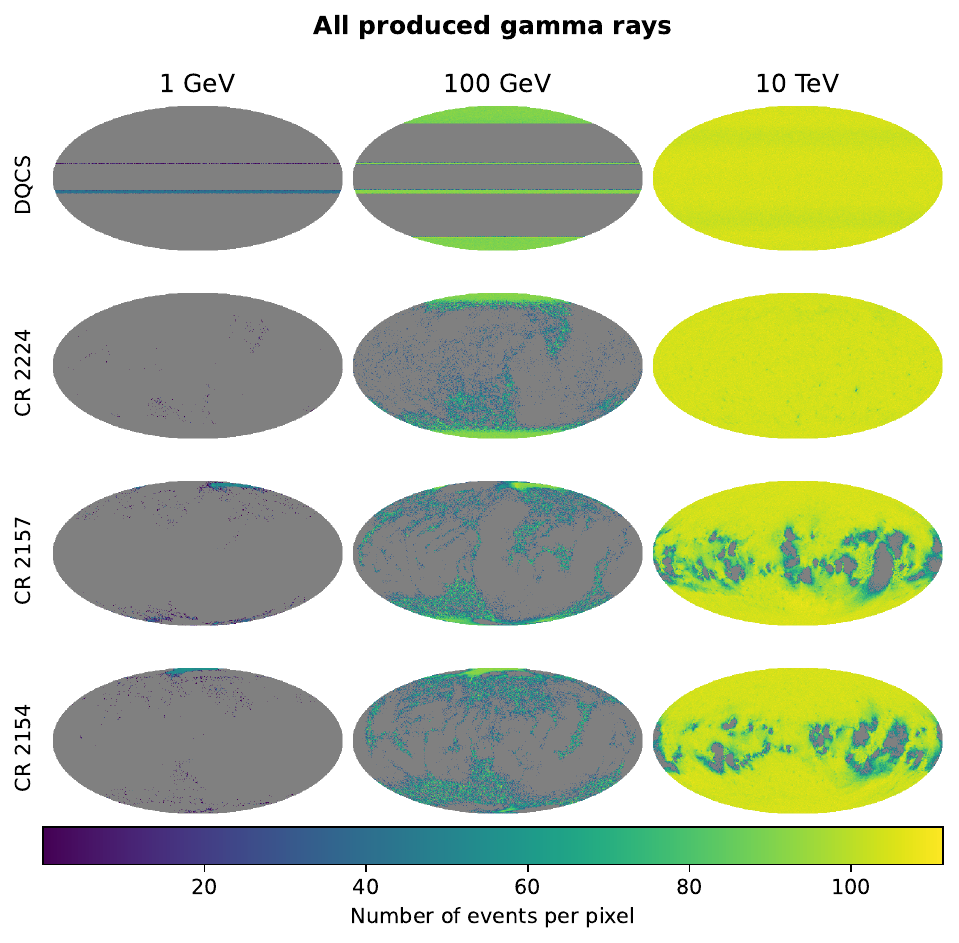}
    \caption{Place of production projected on the solar surface. Here, all produced gamma-rays are shown. The map is in HEALPix format with \texttt{nSide=128}.}
    \label{fig:maps_all}
\end{figure*}

In Fig.~\ref{fig:maps_all}, the place of the gamma-ray production is shown for all produced gamma-rays. The lowest energies (left column) show only very localized places at which the gamma-ray can be produced. Only the DQCS model predicts one more pronounced band in the southern hemisphere. It should be noted that the DQCS model has cylindrical symmetry. Therefore, no longitudinal effects are expected. \change{These latitudinal stripes correspond to the open field originating from the intersection of the quadrupole loops (compare Fig.~\ref{fig:fields}).} Also, the total number of events per pixel is quite low. A better description of the positional features would require simulations with higher statistics, which we leave for a dedicated study. 

\begin{figure*}[hp]
    \centering
    \includegraphics[width=\textwidth]{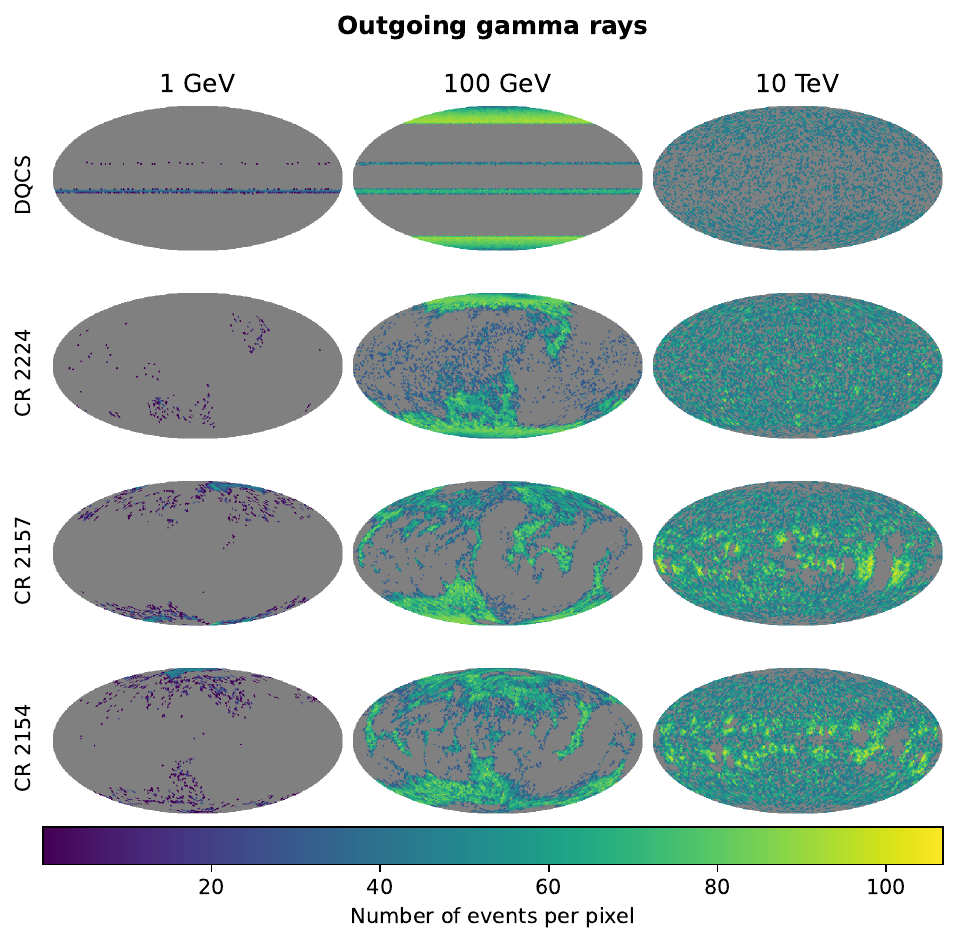}
    \caption{Same as Fig.~\ref{fig:maps_all} for the outgoing photons. Due to the smaller statistics, the resolution is reduced to \texttt{nSide=32} \change{corresponding to a 16 times larger pixel size}.}
    \label{fig:maps_out}
\end{figure*}

\begin{figure*}[hp]
    \centering
    \includegraphics[width=\textwidth]{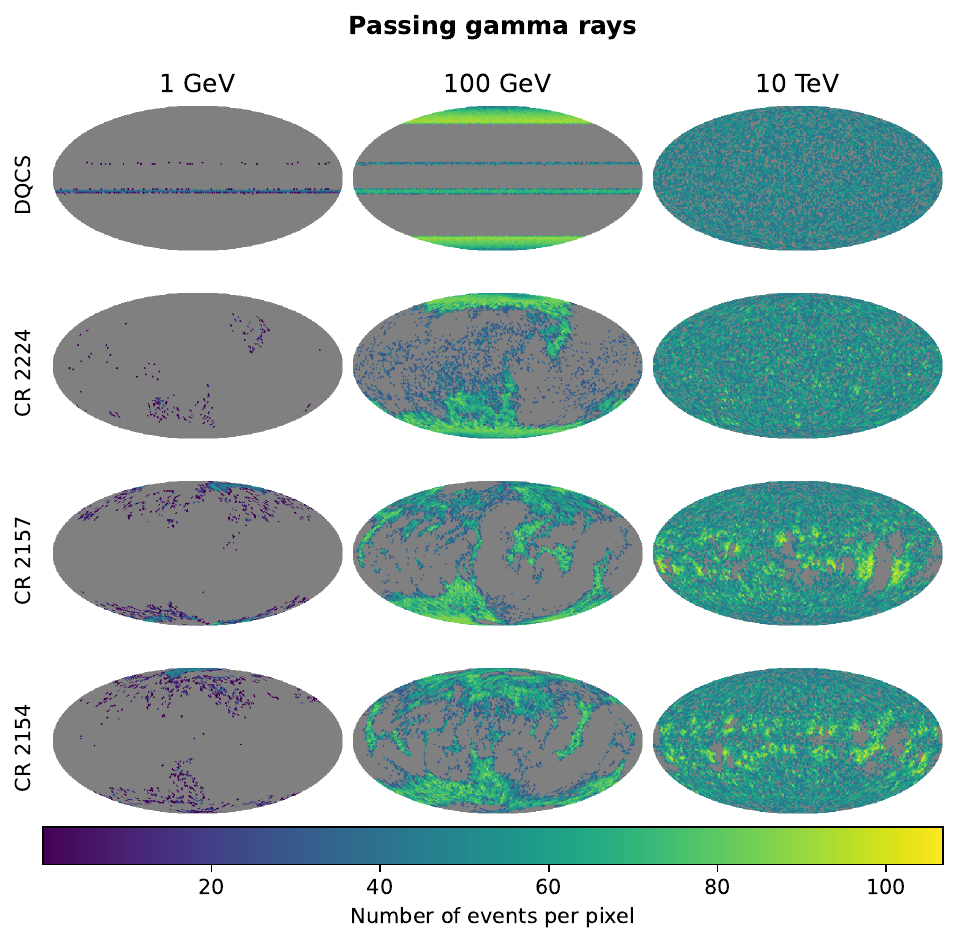}
    \caption{Same as Fig.~\ref{fig:maps_out} for the passing photons.}
    \label{fig:maps_pas}
\end{figure*}

In the intermediate energy (middle column of Fig.~\ref{fig:maps_all}) more substructures can be identified. In all magnetic field geometries, the solar poles show a strongly enhanced gamma-ray production. This is caused by the open field lines originating from the poles, which directly connect the lower layers of the atmosphere to the interplanetary medium. The GCRs arriving at the outer boundary of the corona at $r = 2.5 \, R_\odot$ are bound to these field lines and end in the polar caps. 
The DQCS field shows two bands at $|b|\approx 15^\circ$, originating from the quadrupole component of the field. The band in the southern hemisphere is stronger due to the impact of the current sheet. The solar minimum case of the PFSS models (CR 2224) shows a similar behaviour at the poles, but the edges are not as sharp compared to the DQCS case. In the central part of the solar disk, some more extended features can be seen. The more active cases of CR 2154 and 2157 show smaller hotspots at the polar caps connected to coronal holes. 
Also, the structures in the central disk are more filamentary and separated. While some larger structures, especially in the southern hemisphere, stay between the CR 2154 and 2157, most of the smaller features are uncorrelated. 
These behaviours are expected from the evolution over an 11-year solar cycle of coronal holes, where magnetic field lines are primarily open. During the solar minimum, the polar coronal holes are large and dominate the polar caps, while low-latitude coronal holes are minimal or even non-existent. During the solar maximum, the polar coronal holes shrink and eventually disappear at the peak of the cycle as the Sun’s global magnetic field reverses its polarities, while low-latitude coronal holes are numerous and evolve rapidly.
This shows the critical role of varying solar magnetic field configuration in the solar disk gamma-ray production by the GCRs. 

At the highest energies (right column of Fig.~\ref{fig:maps_all}), the gamma-ray production is distributed homogeneously over the full solar disk in the case of the ordered magnetic field configuration (DQCS model and CR 2224). In the case of the more chaotic field geometries, the impact of the active regions can be identified. Those regions of higher magnetic fields on the surface appear as voids in the gamma-ray production. This is a combination of two effects. On the one hand, the higher magnetic field strength shifts the point of mirroring towards the higher layers of the atmosphere, and the GCRs do not traverse sufficient target material to trigger interactions. On the other hand, the open field lines in the center are pushed towards the edge of the active region, and the GCRs are following them. The total number of produced gamma rays per bin is higher as the number of secondaries per interaction increases with energy. The difference to the more structured appearance at the lower energies, GCRs at the highest energies have a gyro-radius that is large enough to average over most of the field structure. This allows GCRs to change from open to closed field lines and reach the solar surface at nearly all places.

In Fig.~\ref{fig:maps_out} and Fig.~\ref{fig:maps_pas}, the maps of the production places are shown for the outgoing and passing event selection (see Sec.~\ref{ssec:directions} for explanation). Both selections show the same spatial distribution. Therefore, we discuss these maps together and show them for completeness.
In all cases, the total statistic is much lower. Therefore, the maps are converted to a lower angular resolution (\texttt{nSide=32}). 

The general structure aligns well with the maps for all produced gamma-rays. The main difference is the intensity distribution within local features. In the DQCS field geometry with the intermediate energy, the gamma-ray production is shifted towards the edges of the polar cap regions, where it was uniformly distributed in all gamma-ray production. This corresponds to the strong magnetic field at these transition points, which enhances the magnetic mirroring and therefore the ratio between the number of outgoing and total produced gamma rays. In the other field models at the lower and intermediate energies, this effect is not visible as the size of the structure is too small and the edges can not be identified. 

The same effect appears at the edges of the active regions in the high-energy simulation for the active Sun (CR 2154 and 2157). Here, the region around the solar equator shows an enhanced gamma-ray production, as most of the active regions are located here. 
In the case of the more ordered field geometries, the mostly uniform distribution of gamma-rays is conserved, but due to the limited statistics, the simulation results are noisier. 

\begin{figure}
    \centering
    \includegraphics[width=\linewidth]{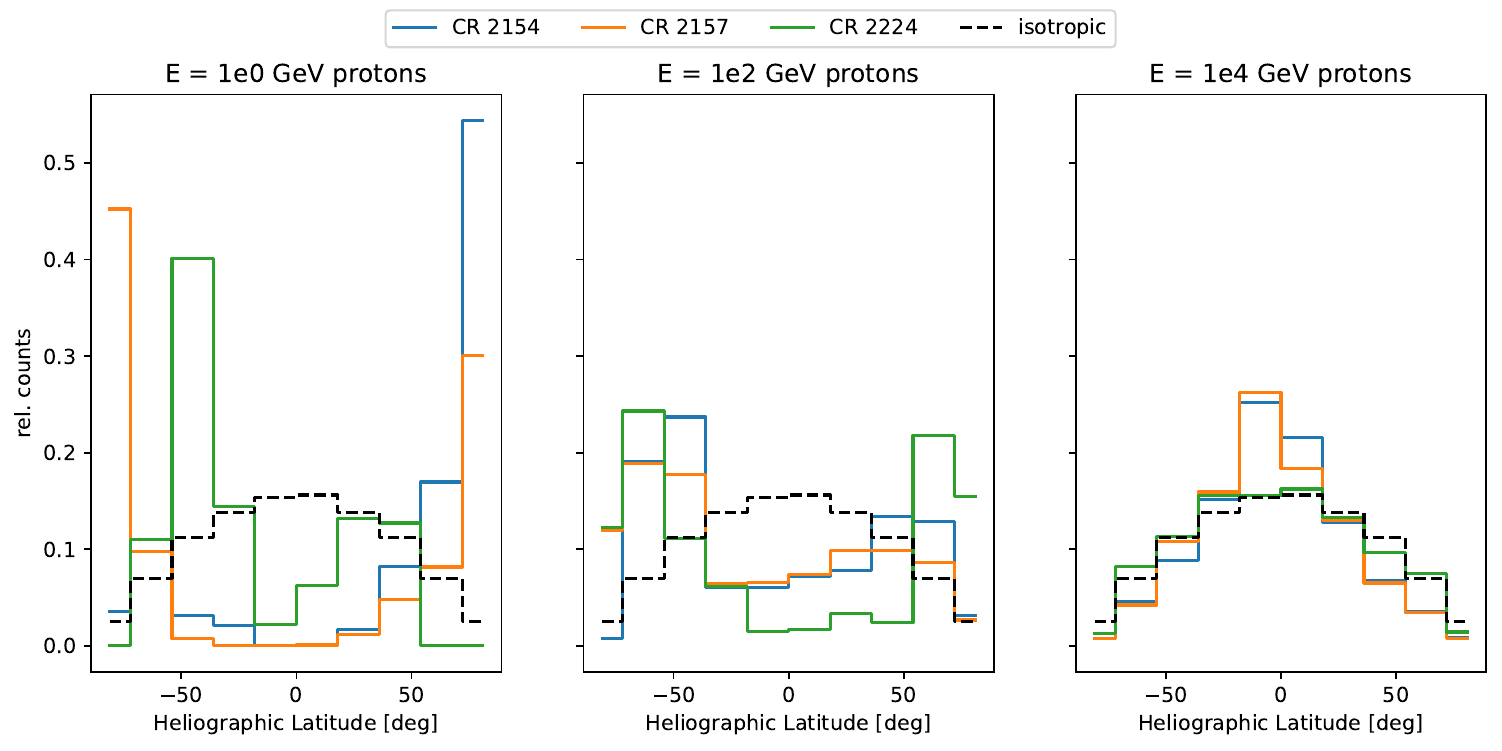}
    \caption{\change{Latitudinal profiles of the gamma-ray distribution calculated from the map of passing photons shown in Fig.~\ref{fig:maps_pas}.}}
    \label{fig:profile}
\end{figure}

\change{With current instruments an observation of the two dimensional distribution of the gamma-rays on the solar surface is not feasible. Therefore, we compute latitudinal profiles similar to the analysis presented in \citet{Linden-etal-2018}. Here, we restrict the analysis only to the surface maps from the passing event selection displayed in Fig.~\ref{fig:maps_pas}, as this is the most realistic one to be compared to observation. The resulting profiles are shown in Fig.~\ref{fig:profile}. Our simulation are performed for mono-energetic GCR protons and consider photons of all energies. Therefore, a direct comparison to the observation by \citet{Linden-etal-2018} is not possible. Possible tensions between our prediction and the observations, have to been investigated further using more realistic magnetic field models and higher statistics observation.}

Overall, the spatial distribution of the gamma-ray production is a promising observable to identify the impact of the solar magnetic field. The effect is strongly energy dependent, which opens the opportunity to combine results from the direct space-based observations at lower energies with the indirect observations with air-shower detectors at higher energies to get the best overview. Nevertheless, future observations need a better angular resolution to differentiate the position on the solar surface, which current generation observatories do not provide.

\section{Realistic injection} \label{sec:energyInjection}

After the mono-energetic injection discussed above, we present the results from the more realistic source injection. In this scenario, the injection spectrum is assumed to follow the local interstellar spectrum (LIS) parametrized in Eq.~\ref{eq:LIS}. At the beginning, we assume the LIS to be isotropic at $r = 2.5R_\odot$. 

\subsection{Direction dependent spectrum} \label{ssec:SED_direction}
\begin{figure*}[ht]
    \centering
    \includegraphics[width=.75\linewidth]{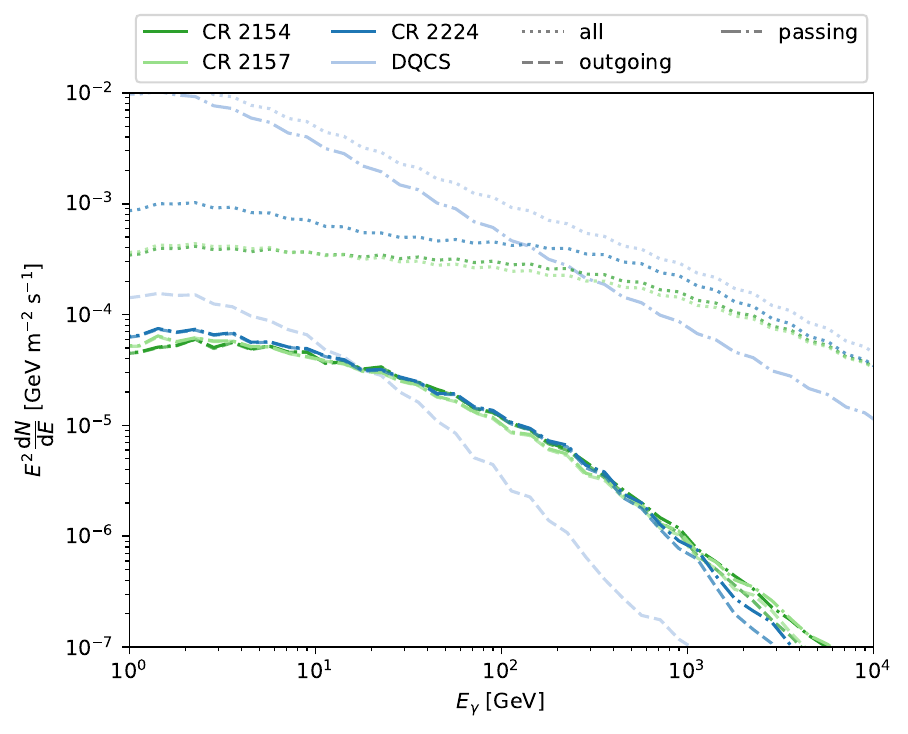}
    \caption{Spectral energy distribution of the produced gamma rays based on a realistic injection spectrum. Different line-styles denote the event selection, and the color shows the magnetic field model.}
    \label{fig:SED}
\end{figure*}

In the first step we shows the resulting Spectral Energy Distribution (SED) based on the different event selections discussed in Sec.~\ref{ssec:directions} and magnetic field configurations in Fig.~\ref{fig:SED}. 
While the PFSS models for different Carrington Rotations show a similar behavior, the energy spectrum predicted in the DQCS magnetic field configuration deviates strongly. 
The selection of all created gamma-rays and all gamma-rays that pass the surface does not show any break in the SED, and the spectral slope relates directly to the incoming proton SED. Only the selection of outgoing gamma-rays, which require a magnetic mirroring of the parent proton, shows a change in the spectral slope at $\sim 10 \, \mathrm{GeV}$. 

The PFSS models predict similar SEDs for all Carrington Rotations. In the case of all produced gamma-rays, the highest energies relate directly to the injected proton spectrum, and the normalization of the predicted gamma-ray flux is the same. At lower energies $E_\gamma \lesssim 1 \, \mathrm{TeV}$ the spectral slope changes and becomes harder. Also, the absolute normalization for the solar minimum simulation (CR 2224) is higher than the solar maximum cases (CR 2154 and 2157). 

The selection of outgoing and passing gamma-rays predicts nearly the same flux. Only at the highest energies, small deviations are visible, but those can still be caused by limited statistics. Here, the energy dependence at the highest energies is softer than the injected proton spectrum from the LIS. This shows the additional energy-dependent effect of the magnetic mirroring. The spectrum shows a break at several hundred GeV, and the energy dependence is harder below the break. This behavior is expected from the observations that show a break between 300 and 500 GeV. The difference in absolute normalization between solar minimum and maximum vanishes for the observable gamma-ray flux. 

\subsection{Data comparison} \label{ssec:SED_data}
\begin{figure*}[ht]
    \centering
    \includegraphics[width=.75\linewidth]{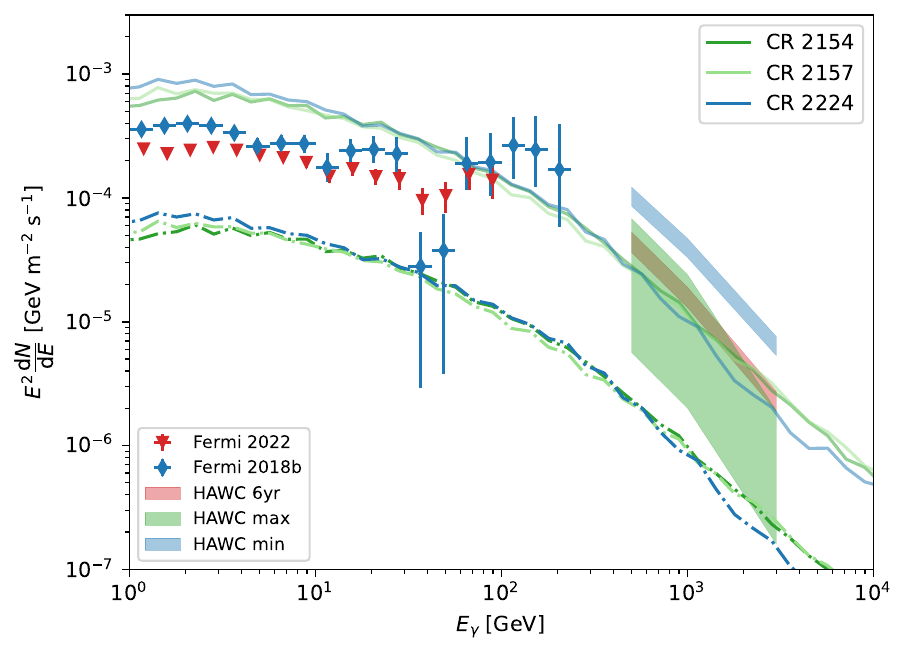}
    \caption{Spectral energy distribution of the passing gamma-rays (dash-dotted lines) compared to the observations with Fermi \citep{Tang-etal-2018,Linden-etal-2022} and HAWC \citep{Albert-etal-2023}. The solid lines show the same simulations scaled by a factor of 12 to include missing contributions for heavier nuclei, the interplanetary magnetic field and the structures in the gas distribution (see Sec.~\ref{ssec:SED_data}).}
    \label{fig:SED_passing}
\end{figure*}

In Fig.~\ref{fig:SED_passing}, the predicted SED in the passing event selection is compared to some observations with the FERMI satellite and with HAWC. The total normalization of the prediction is about one order of magnitude too low compared to observations. At the highest energies, the expected energy scaling $\sim E^{-3.6}$ is reproduced, and the turning towards the harder energy dependence below 300 GeV is reproduced. In the model presented here, no modulation effect has been taken into account. This will affect especially the lowest energetic protons and further suppress the gamma-ray flux at the one to ten GeV scale. 

Additional effects will increase the expected gamma-ray flux. In the following, we discuss the three most important: 
\begin{enumerate}
    \item GCRs do not only consist of protons but also heavier elements. \cite{Castro2025} discusses the impact of heavier elements on the total diffuse gamma-ray flux. 
    \change{These authors use measurements for the contribution by GCR proton and helium and estimate limits for a heavier component by assuming a carbon or iron dominance.}
    They conclude that the total flux can be increased by a factor of up to 2 compared to a proton-only model. 

    \item For the initial pitch-angle distribution, we assume isotropic momentum at $r = 2.5 R_\odot$. This is only valid for strong scattering of particles reaching this outer boundary of the solar atmosphere. \cite{Hutchinson-etal-2022} introduces the effect of magnetic mirroring that can already happen in the Parker magnetic field between the Earth and Sun. Without additional scattering, this will change the pitch-angle distribution. In Appendix \ref{app:mu0} we investigate the effect of a different pitch-angle distribution leading to an increased gamma-ray flux up to a factor 2. Depending on the details of the scattering, i.e., the pitch-angle diffusion coefficient, this effect might become even stronger. 

    \item In this study, we focus on the impact of solar magnetic field configuration. Therefore, we apply a simplified spherical symmetric gas distribution. In nature, one would expect a coupling between the magnetic field structure and the gas distribution. 
    \change{State-of-the-art 3D MHD models, by solving the coupled plasma–magnetic system, can provide physically self-consistent 3D densities that vary with magnetic topology and dynamics, and may yield higher densities in specific regions than a 1D model.}
    This will increase the total gamma-ray emission as the particles will see more column density and the interactions become more likely \citep{Linker-etal-1999, vanDerHolst-etal-2014, Perri-etal-2022}. 

    Besides the increased interactions within the chromosphere, the GCRs reaching the lower boundary of our simulation, the photosphere, can interact in the deeper layers of the Sun \citep{Li-etal-2024}. The combination of both effects can increase the total gamma-ray flux easily by a factor 3 or more. 
\end{enumerate}

To account for the additional flux, we scale the predictions from our model in Fig.~\ref{fig:SED_passing} by a factor of 
12 and achieve a reasonable agreement with the measurements. At the lowest gamma-ray energies, the influence of the solar modulation\change{, which is not included in the model presented here,} is needed to avoid an overshooting of the Fermi measurements. 

\subsection{Expected Neutrino signal}
Hadronic interactions do not only produce high-energy gamma rays, but also create neutrinos as a decay product of charged pions. In contrast to the gamma rays, neutrinos are weakly interacting and can therefore traverse the Sun. In a realistic description, the creation and propagation of neutrinos is necessary. Up to now, no evidence for high-energy neutrinos originating from the Sun has been found \citep{IceCube-2025}. Therefore, we start here with a simplified estimate of the expected neutrino flux from the Sun. 

To convert our simulated gamma-ray flux into a neutrino signal, we follow the so-called \textit{delta-approximation} \citep[see][and references therein]{Becker-Tjus-Merten-2020}. Positive, negative, and uncharged pions are created in equal manner, leading to a ratio of 1:3 between gamma rays and neutrinos. Here, we assume that all neutrino species have the same abundance at Earth after oscillation. The average energy of a produced neutrino is half the average energy of a gamma ray. Therefore, the total expected neutrino flux can be calculated as 
\begin{equation}
    \Phi_\nu(E) = 3 \cdot \Phi_\gamma(2 E) \quad .
\end{equation}

The flux of the neutrinos reaching the Earth can be divided into two components. The first component is connected to the gamma-ray production in the outgoing or passing direction. Those neutrinos are directed away from the Sun, and no absorption effect has to be taken into account. The second contribution are the \textit{behind the limb} events. Here, the GCR interaction happens in the direction of the solar disc. While the corresponding gamma-rays are absorbed and can not be observed, some of the neutrinos can still travel through the Sun. 
To calculate the fraction of neutrinos passing the Sun, we assume the neutrino nucleon cross section presented by \citet{Formaggio-Zeller-2012} and a homogeneously filled sphere. 

\begin{figure*}[ht]
    \centering
    \includegraphics[width=.8\linewidth]{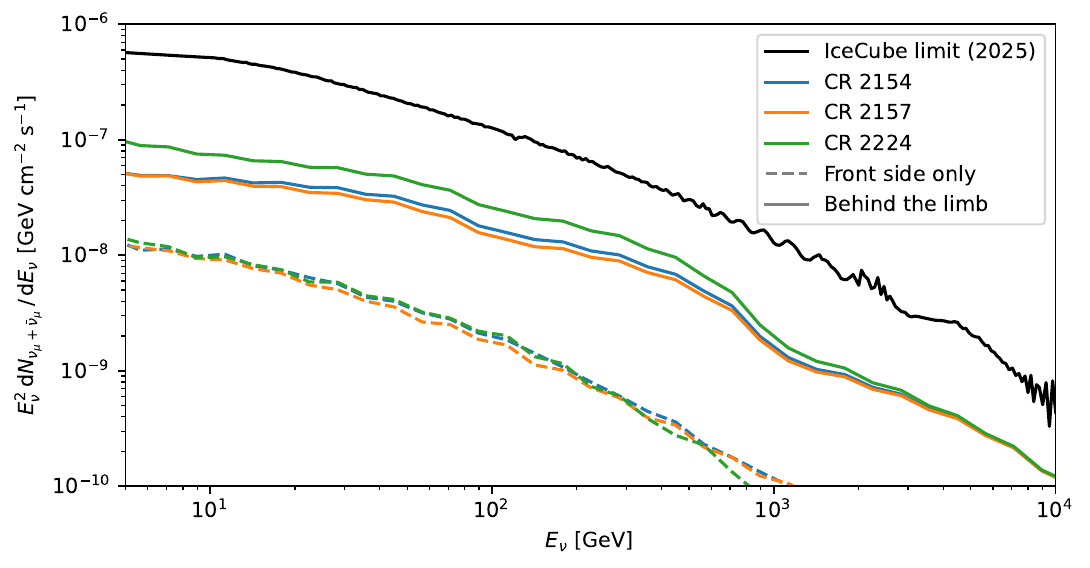}
    \caption{Expected muon neutrino flux from the hadronic interactions in the solar atmosphere. The IceCube upper limit is taken from \cite{IceCube-2025}.}
    \label{fig:neutrinos}
\end{figure*}

The resulting estimate for the muon neutrino flux is shown in Fig.~\ref{fig:neutrinos} together with the upper limit from the IceCube neutrino observatory \citep{IceCube-2025}. The total neutrino flux is dominated by the \textit{behind the limb} events. The spectral shape is similar to the model-based upper limit derived by IceCube. The normalization is a factor $\sim 5$ lower than the limit. This is a comparable value to the IceCube results, where the limit is $2.48$ times higher than the tested model. Overall, a neutrino detection seems possible in the near future with more data and an improved detector for the IceCube upgrade.

\section{Discussion}
Accurate modeling of the hadronic solar gamma-ray emission with detailed spectral and spatial information remains a challenging task. Our simulations demonstrate that the large-scale solar magnetic field exerts a significant influence on both the spectral properties and the morphology of gamma-ray emission from the solar disk. The dependence of the emission on pitch-angle distributions is clearly demonstrated within our model framework. While our baseline assumption of isotropy at 2.5 $R_{\odot}$ provides a useful reference case, it is clear that the Parker spiral modifies the distribution of incoming cosmic rays well before they reach the near-Sun region. Magnetic mirroring in the heliospheric field between 1 AU and a few solar radii has been shown to preferentially select forward-directed particles, thereby potentially modifying the gamma-ray yield by allowing more of the cosmic rays with such a pitch-angle distribution to penetrate to deeper layers of the chromosphere \citep{Hutchinson-etal-2022}. This effect, together with uncertainties in the scattering conditions of the inner heliosphere, suggests that transport physics remains a key open problem for interpreting solar gamma-ray observations.

Recent studies have emphasized the role of small-scale magnetic structures at the solar surface in shaping gamma-ray emission. \citet{Li-etal-2024, Li-etal-2025} demonstrated that flux-tubes and intergranular sheets can significantly increase the reflection probability of GCRs, thereby boosting the gamma-ray yield. Our results similarly support the significant impact of localized structuring, especially near active region boundaries and coronal holes. The discrepancy between our predictions and observations thus points to the need for multi-scale models that combine global coronal fields with chromospheric and photospheric fine structure, especially in target density variations.
Future observations, i.e., in the multi TeV range, will help to distinguish between different production locations. While \citet{Li-etal-2025} predict an exponential cut-off at a few TeV for the gamma rays originating in the internetwork regions, the power-law in our model expands towards the $\sim 10$ TeV region.

Solar modulation of the GCR spectra arriving near the Sun should only be of relevance for relatively low energies. For GeV range GCRs and solar gamma rays, it may be of interest to investigate the influence of the modulation effect on resulting gamma-ray spectra and a potential temporal signature related to the solar cycle \citep[e.g.,][]{Acharyya-etal-2025}. Additionally, transient signatures in the GeV range are also of significant interest due to Fermi observations in recent years \citep{Ackermann-etal-2014,Ajello-2021}. The primary ions causing the gamma-ray emission could originate directly from solar flare eruptions \citep[e.g.,][]{Ryan-Lee-1991, Ryan-2000, Hurford-etal-2006, deNolfo-etal-2019, Pesce-Rollins-2024, Bruno-etal-2025}, coronal mass ejection shock waves \citep{Cliver-etal-1993,Pesce-Rollins-etal-2015,Ackermann-etal-2017,Jin-2018, Petrosian-2018,Gopalswamy-etal-2018, Kahler-etal-2018, Makela-etal-2023} or additional transient effects such as the coronal shock wave during eruptions \citep{Warmuth-2025}. In all such scenarios, the transport effects from the origin of the primary particles to the gamma-ray emission site and the associated field geometries remain a crucial factor. Our CRPropa-based model presented here is well equipped to address these questions in future studies with comprehensive end-to-end modelling.

The coronal field models employed in this study, namely the PFSS model, while relatively simple to determine from magnetic maps, are limited in their capabilities to accurately describe the smaller scale structures of the solar field. While improvements to the PFSS field approach are possible \citep[e.g.\ by modifying to an elliptical source surface][]{Kruse-2021}, more sophisticated models based on non-linear force-free extrapolations \citep{Wiegelmann-2004} or inputs directly from MHD simulations \citep{Jin-etal-2016} should be studied in the future.

\begin{acknowledgments}
 HF and JBT acknowledge funding from the German Science Foundation DFG project "Modeling the Time-Dependent Cosmic-Ray Sun Shadow and its related Gamma-Ray and Neutrino Signatures" (FI~706/27-1 and TJ~62/7-1, project number 437789084). JD, FE, HF, JBT acknowledge further funding from the DFG via the Collaborative Research Center SFB1491 "Cosmic Interacting Matters - From Source to Signal" (project number~445990517).
MJ, WL, and VP acknowledge support by NASA LWS grant 80NSSC21K1327.
\end{acknowledgments}

\bibliography{references}

\begin{appendix}
    
\section{Influence of the Magnetic Mirror in the Parker Field} \label{app:mu0}

In this section, we investigate the influence of the pitch-angle distribution (PAD) on the gamma-ray production. In the main part, we discussed an uniform PAD at the outer surface ($r = 2.5R_\odot$). This assumption is valid in the case where the mean free path of the scattering $\lambda_\mathrm{mfp}$ is much shorter than the distance between Earth and Sun. 
\cite{Hutchinson-etal-2022} introduce the case of weak scattering, where the magnetic mirroring of GCRs within the Parker magnetic field becomes relevant. In the following, we introduce the PAD resulting from the mirroring in the Parker field (Sec.~\ref{app:PAD}) and estimate the changes in the gamma-ray flux based on the new PAD (Sec.~\ref{app:Gamma-flux}). 

\subsection{Pitch-angle distribution from mirroring} \label{app:PAD}

To estimate the PAD of particles undergoing the magnetic mirroring, the trajectory of $10^8$ particles in the Parker magnetic field
\begin{equation}
    \vec{B}(\vec{r}) = B_0 \left(\frac{r}{r_0}\right)^{-2} \vec{e}_r  - B_0 \frac{r_0^2 \Omega \sin\theta}{v_\mathrm{w} r} \vec{e}_\phi 
\end{equation}
is calculated. The reference field strength $B_0 = 187 \, \mathrm{\mu T}$ at the reference radius $r_0 = R_\odot$ is chosen to match the observed field strength at the Earth. The solar wind speed is set to $v_\mathrm{w} = 500 \, \mathrm{km/s}$ and the rotation is $\Omega = 2.86 \cdot 10^{-6} \, \mathrm{rad/s}$. 
All particles are injected on a sphere at $r = 1\, \mathrm{au}$ with an isotropic injection. The particles are observed at three different radial positions ($r/R_\odot = 2.5, 5, 7.5$) and all particles reaching the inner boundary are lost. 

\begin{figure}[t]
    \centering
    \includegraphics[width=0.6\linewidth]{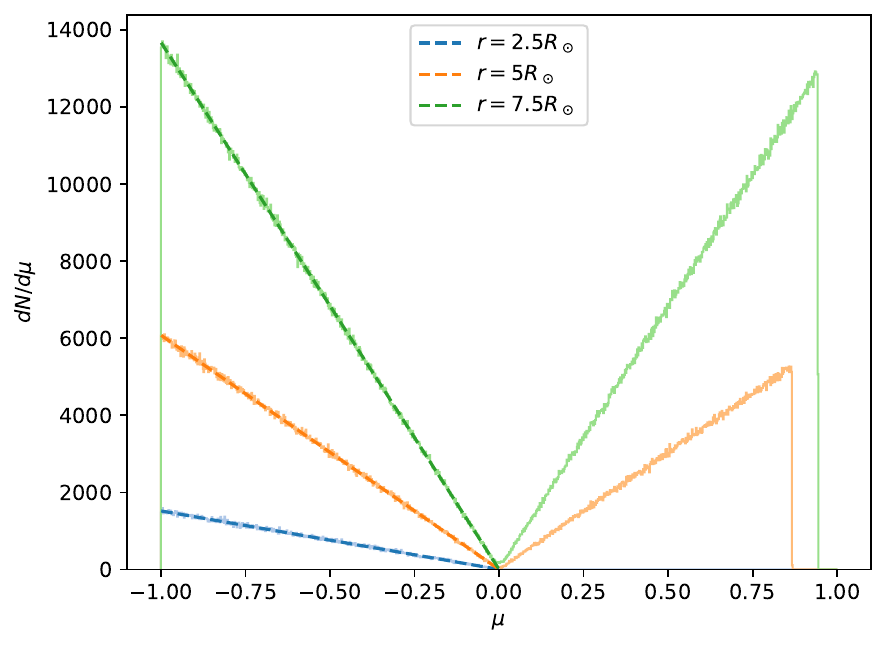}
    \caption{Pitch-angle distribution for particles at different distances from the Sun. The particles are injected at $r = 1\, \mathrm{au}$ with an isotropic momentum distribution.}
    \label{fig:PAD}
\end{figure}

In Fig.~\ref{fig:PAD}, the PAD of the observed particles is shown. The distribution of the ingoing particles follows a linear distribution. For the larger distances ($r = 5 R_\odot$ and $r=7.5 R_\odot$) the mirrored particles that are leaving the sphere can be seen. 
The PAD of particles in the loss-cone of the magnetic mirror can be described by 
\begin{equation}
    \left.\frac{\mathrm{d} N}{\mathrm{d} \mu}\right|_\mathrm{MM} = - 2 \mu \quad. 
\end{equation}
The negative sign considers only ingoing particles and the total probability is normalized $\int_{-1}^{0}\mathrm{d}\mu \, \frac{ \mathrm{d}N}{\mathrm{d}\mu} = 1$. 

\subsection{Gamma-ray flux} \label{app:Gamma-flux}

In the next step, the PAD from the magnetic mirroring is applied to the calculation of the gamma-ray SED. Here, we use the dataset simulated before and extend the re-weighting to account for the initial PAD. The weights from Eq.~\ref{eq:weights} are extended by the factor for the initial pitch-angle. The full applied weight reads as: 
\begin{equation}
    w_i^{\mathrm{(full)}} = w_i \cdot \frac{\left. \frac{\mathrm{d}N}{\mathrm{d}\mu}\right|_\mathrm{MM}}{\left. \frac{\mathrm{d}N}{\mathrm{d}\mu}\right|_\mathrm{sim}} \quad,
\end{equation}
where the simulated PAD is flat $\frac{\mathrm{d}N}{\mathrm{d}\mu}\big|_\mathrm{sim} = 1$. 

\begin{figure}[ht]
    \centering
    \includegraphics[width=.6\linewidth]{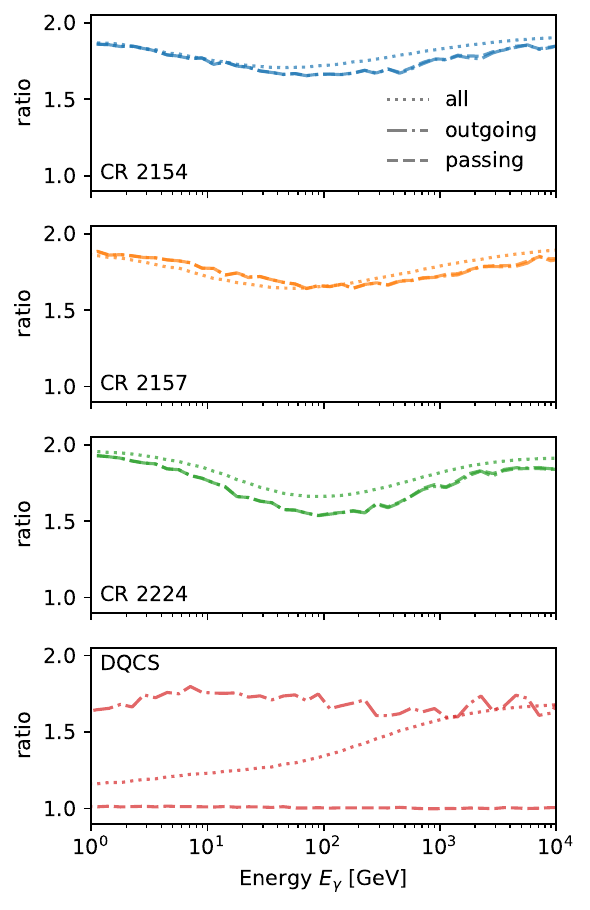}
    \caption{Ratio between the gamma-ray flux from the mirrored PAD to the uniform PAD}
    \label{fig:SED_ratio_mu0}
\end{figure}

From the re-weighted flux, we calculate the ratio between the magnetic mirror PAD to the isotropic scenario. The result is shown in Fig.~\ref{fig:SED_ratio_mu0}. The cases of the PFSS magnetic field configuration show a clear increase of the total gamma-ray flux between 70\% and 90 \%. The dip around $E_\gamma \approx 100 \, \mathrm{GeV}$ indicates a shift in the break energy. The total increase is expected as more particles start with a strong forward direction, and can penetrate deeper into the atmosphere and undergo more interactions. 

The case of the DQCS magnetic field shows more complex behavior depending on the selected events. The outgoing gamma-rays are increased by roughly 70\%, as more particles are in the forward direction and will be mirrored deeper within the solar atmosphere. The passing events are nearly unaffected by the changes in the PAD. This indicates that these gamma-rays are mainly caused by GCRs following the current sheet into the deeper layer of the atmosphere. This effect does not depend on the initial pitch angle. The line of all produced gamma-rays shows an energy-dependent transition between these two cases.  

\end{appendix}
\end{document}